\def\preprint{0}		
\def\preprint{1}		

\if\preprint1
	\documentclass{article}
	\usepackage{mn2e}
	\usepackage{astrop-bib}
	\usepackage{times}
	\usepackage{graphicx}
	\usepackage{calc}

\else
	\documentstyle[astrop-bib,referee,times]{mn}
	\newcommand{\includegraphics}[1]{}
\fi

\title[Wavelength-dependent angular diameters: an observational perspective]{Wavelength dependence of angular diameters of M giants: an observational perspective \protect \\}

\author[A.~P.~Jacob et al.]
        {A.~P.~Jacob,$^1$\thanks{email: \tt ande@physics.usyd.edu.au}
        T.~R.~Bedding,$^1$
	J.~G.~Robertson,$^1$
        M.~Scholz$^2$\\
        $^1$School of Physics, University of Sydney 2006, Australia\\
	$^2$Institut f. Theoretische Astrophysik der Universit{\"a}t Heidelberg, Tiergartenstr. 15, 69121 Heidelberg, Germany\\}

\date{Accepted ????.
      Received ????;
      in original form ????}
\pubyear{1999}

\begin{document}

\maketitle

\begin{abstract}
We discuss the wavelength dependence of angular diameters of M giants from an observational perspective. Observers cannot directly measure an optical-depth radius for a star, despite this being a common theoretical definition. Instead, they can use an interferometer to measure the square of the fringe visibility. We present new plots of the wavelength-dependent centre-to-limb variation (CLV) of intensity of the stellar disk as well as visibility for Mira and non-Mira M giant models. We use the terms ``CLV spectra'' and ``visibility spectra'' for these plots. We discuss a model-predicted extreme limb-darkening effect (also called the narrow-bright-core effect) in very strong TiO bands which can lead to a misinterpretation of the size of a star in these bands. We find no evidence as yet that this effect occurs in real stars. Our CLV spectra can explain the similarity in visibilities of R~Dor (M8IIIe) that have been observed recently despite the use of two different passbands. We compare several observations with models and find the models generally under-estimate the observed variation in visibility with wavelength. We present CLV and visibility spectra for a model that is applicable to the M supergiant $\alpha$~Ori.

\end{abstract}


\section{Introduction}
The radius of a star is considered to be one of its fundamental properties. Stellar radii may be determined from interferometric observations of stellar angular diameters, but stars do not necessarily appear to us as sharp-edged disks of uniform brightness and size. Rather, two effects may be present: limb-darkening which tapers the surface brightness down towards the star's edge and line absorption which changes the star's apparent size. Both effects depend on the wavelength of the observation and are particularly significant in M giants.

To determine the radii of M giants from interferometric data, observers have relied upon comparisons with models of the centre-to-limb variation (CLV) of the intensity of the stellar disk. The simplest CLV, a uniform disk, is often used for this. Alternatively we can use CLVs derived from appropriate model photospheres of late M giants, since this procedure can lead to improvements in these models and a greater understanding of these stars. The model-photosphere CLVs are defined for a particular filter or spectral passband and must be transformed to visibilities for comparison with the observations. Model CLVs for late M giants were first studied by \citeone{WK79} and \citename{ST87} (\citeyear{ST87}; hereafter ST87), and were recently presented for non-Miras by \citename{HS98} (\citeyear{HS98}; hereafter HS98) and for Miras by \citename{HSW98} (\citeyear{HSW98}; hereafter HSW98).

Model CLVs from these studies have been used for comparison with observations of M giants (for example, \citeone{HST95}, \citeone{WBH96} and \citeone{PCR99}). \citeone{HST95} have shown that, in some cases at least, Gaussian CLVs are better fits to the observations than the model CLVs. Other work, e.g. \citeone{WBB92}, has shown that some M giants do not appear circular and some show asymmetrical brightness distributions. These effects have not yet been included in any model. Most observations to date have been made at only a few isolated, narrow passbands that were determined by the availability of filters, the need to prevent fringe-smearing and the position of the deep near-IR absorption features in M giant spectra. Recently \citeone{BBB98} and \citeone{YBB99} systematically studied the variation in angular diameter with time of the Miras R~Leo and $\chi$~Cyg and showed that periodic diameter variations did occur.

To further improve the M giant models, \citeone{S85} and \citeone{BBSW89b} have called for measurements of M giant radii over a wide range of wavelengths. Fitting wavelength-dependent model CLVs to multi-wavelength interferometric observations, at a single epoch, should be a more robust process for comparing models and observations than simply fitting a small number of model CLVs to narrow-band observations. This should provide a useful tool for improving our understanding of M giants, particularly their temperature structure.

In this paper we present the first stage of this work. We calculate wavelength-dependent CLVs for several models and use them to derive wavelength-dependent visibility profiles. We discuss how these visibility profiles can be compared with observations. We adopt the perspective of an observer who cannot measure an optical-depth radius (the usual theoretical radius definition) and has limited spatial frequency coverage and therefore limited accuracy in reconstructing the true CLV. We discuss an extreme limb-darkening effect that occurs in some models and compare some past observations with wavelength-dependent visibilities derived from appropriate models. As an example, we consider the star R~Doradus (M8IIIe) which is a semi-regular variable with Mira-like properties (\citename{BZJ98} 1997, 1998). Interferometric observations are usually carried out through narrow-band interference filters to avoid fringe smearing effects but we have acquired a large set of multi-wavelength observations of R~Dor using MAPPIT (Masked APerture-Plane Interference Telescope, see \citebare{BRM94}, \citebare{JBR97}) which are currently being processed. For these observations MAPPIT included a prism to disperse the interference fringes, thus producing multi-wavelength interferometric observations.

\section{Stellar Radii}
Although the stellar radius, $R$, and its associated effective temperature $T_{\rm eff} \propto (L/R^2)^{1/4}$ are convenient and useful parameters for describing a star's global properties, limb-darkening and line absorption effects mean we have to choose an outer layer whose distance from the centre of the star we {\it define} to be the radius. Baschek et al. (1991) have discussed various radius definitions occurring in the literature. The difficulties in formulating meaningful definitions are most severe for stars whose outer portions have low density gradients and are strongly extended, e.g., Wolf-Rayet stars and Mira variables. These stars have extended atmospheres whose geometrical thickness is not small compared to the total dimensions of the stars, in contrast to the compact or plane-stratified atmospheres of the vast majority of stars in which almost all photons seen by an observer originate at essentially the same distance from the star's centre. But even main sequence stars such as the Sun present ambiguities, as shown by the difference between the photospheric, coronal and radio sizes of the solar disk.

Only for the Sun can we easily measure the brightness distribution of the disk. For other stars we can use an interferometer to measure the (square of the) fringe visibility generated by the brightness distribution. Since the visibility is usually only measured out to, or just past, the first null and with errors of several percent, one cannot uniquely reconstruct the CLV. In some cases a null is not present and we are limited simply by the available spatial frequency coverage. One can, at best, exclude certain extreme CLVs if the data are accurate enough (see references given in \citebare{S97}; hereafter S97, and \citebare{HAH98}). Instead of reconstructing the CLV, observations are usually fitted with model curves or, if no suitable model is available, with a standard limb-darkening profile such as a uniform disk (UD), fully-darkened disk (FDD) or a Gaussian. A point somewhere along the CLV can then be chosen to be the radius. For the Sun the inflection point is used for defining the photospheric radius. Thus, the radius we assign to a star is dependent on both the model and the chosen radius definition. 

The most common monochromatic radius definition to be found in theoretical work is the position of the $\tau_{\lambda} = 1$ layer, i.e. the radius on a line of sight to the disk's centre at which the optical depth reaches unity for a very narrow passband with central wavelength $\lambda$. However, this is $\it not$ $\it an$ $\it observable$ quantity, even for the Sun, and there is no simple rule that gives the $\tau_{\rm \lambda}=1$ point on the CLV. It must be be taken from a model. This optical-depth radius describes the ``typical'' layer from which photons seen by the observer are emitted. One must be aware that the intensity contribution function covers a wide range of depths around this layer, which may lead, at different wavelengths, to quite different CLVs and quite different $\tau_{\lambda}=1$ positions on these CLVs. We refer to S97, HS98 and HSW98 for details. For broad-band or whole-spectrum definitions of the radius, some method of averaging the optical depth radius over wavelength is required. It is common in this case to use the Rosseland radius, i.e. the radius at which the Rosseland optical depth equals unity ($\tau_{\rm Ross}=1$), as this is often used in modelling studies.

The CLV takes into account both limb-darkening and size variation resulting from (molecular) line absorption effects. The difficulty in assigning a radius to M giants, both Miras and non-Miras, is increased by their strongly extended atmospheres and by the almost complete failure of the simple UD approximation. Instead, model CLVs similar to FDD and Gaussian shapes are more appropriate. In these cases it is usual to define a radius to avoid using the fully tabulated description of the surface brightness distribution for measuring the star's ``size''. On the other hand, since M giants do not have a unique radius it may be more useful to compare the observed visibility profile with the model visibility profile derived from the model CLV. To do this we need to choose an appropriate model.

\section{Model CLVs and Visibilities}
\subsection{Models}
We selected from the grid of \citename{BBSW89a} (\citeyear{BBSW89a}, \citeyear{BBSW91}; hereafter collectively BBSW) the 1M$_{\odot}$ static (i.e. non-Mira) AGB model, referred to as ``X280'' in BBSW. This model's effective temperature and atmospheric extension (measured as the difference between the uniform disk radii in TiO and pseudo-continuum bands, as given in HS98) are close to the observed values of R~Dor. \citeone{BZvdL97} found the effective temperature of that star to be 2740\,K and \citeone{JBR97} estimated the extension to be 20\%. Here we use the term ``pseudo-continuum'' to refer to the high-flux spectral regions between the TiO bands that lie below about 1\,$\mu \rm m$, since the flux here is still well below the true continuum. The BBSW models are static but are suggested by HS98 to be fair approximations to semi-regular variables like R~Dor, whose atmospheric structures are less dominated by outflow and inflow dynamics than those of Mira variables. However, since R~Dor also shows some Mira-like properties, a 1M$_{\odot}$ Mira model, referred to as ``E8380'' in \citename{BSW96} (\citeyear{BSW96}; hereafter BSW96), was also investigated. A further Mira model of the P series (``P74200'') of HSW98 is used for discussing extreme forms of limb-darkening in strong TiO bands. We also investigate a model of the H series (``H350'') of BBSW, which is applicable to the well-studied M supergiant $\alpha$~Ori. See Table~\ref{modelparams} for basic parameters of the models used. Note that although none of the models are tailored specifically for the stars listed in the last column they are representative and, in some cases, have been used in previous studies for these stars. The ``parent-star'' concept for Mira models is explained in BSW96.

We refer to BBSW, BSW96 and HSW98 for an extensive discussion of these static and Mira models and the current approximations used in the treatment of molecular band absorption, which result in uncertainties in the calculated spectrum and CLV in strongly saturated TiO bands. Nevertheless these models are the most sophisticated presently available for these types of stars.

\begin{table*}
\caption{Basic parameters of models: visibility phase, period $P$, mass $M$, luminosity $L$, Rosseland radius of the non-pulsating parent-star of the Mira model series $R_p$, Rosseland radius of non-Mira and Mira model stars $R$ and effective temperature $T_{\rm eff} \propto (L/R_{(\tau_{\rm Ross}=1)}^2)^{1/4}$. The X and H models are static, the P and D models are fundamental mode Miras and the E model is a first-overtone Mira. The last column lists the stars to which the models are compared in this paper. }
\begin{center}
\begin{tabular}{lcccccccccc}
\hline
Model & Vis. phase & $P$/days & $M/M_{\odot}$ & $L/L_{\odot}$ & $R_p/R_{\odot}$ & $R_{(\tau_{\rm Ross}=1)}/R_{\odot}$ & $R_{(\tau_{\rm {Ross}=1})}/R_p$ & $T_{\rm eff}$/K & Reference & Star\\
\hline
X280  &-   &--  &1  &$10^4$ &--  & 426 & --   & 2800 & BBSW  & R~Dor        \\
H350  &-   &--  &15 &$10^5$ &--  & 862 & --   & 3500 & BBSW  & $\alpha$~Ori \\
E8380 &1.00&328 &1  &6750   &366 & --  & 1.09 & 2620 & BSW96 & R~Dor, R~Cas \\
E8560 &1.21&328 &1  &7650   &366 & --  & 1.17 & 2610 & BSW96 & R~Leo        \\
P74200&2.00&332 &1  &4960   &241 & --  & 1.04 & 3060 & HSW98 & -            \\
D28320&1.80&330 &1  &3510   &236 & --  & 0.90 & 3050 & BSW96 & R~Cas        \\
D28760&2.00&330 &1  &4560   &236 & --  & 1.04 & 3030 & BSW96 & R~Cas        \\
D28847&2.20&330 &1  &4760   &236 & --  & 1.09 & 3000 & BSW96 & R~Leo        \\
\hline
\end{tabular}
\end {center}
\label{modelparams}
\end {table*}

\subsection{Model-Derived CLVs}
\label{clvspecplot}
For each model we produced CLVs for wavelengths from 550 to 1100\,nm at 1\,nm intervals. Each CLV was calculated for a rectangular passband of 1\,nm width. In order to define a reference radius for each of these passbands we carried out an intensity-weighted integration of the $\tau_{\rm \lambda}=1$ radii over 1\,nm in wavelength using the filter radius ($\tau_{\rm filter}=1$) definition proposed by ST87. For the purposes of this paper, we will refer to these narrow passbands, their CLVs and filter radii as monochromatic.

\begin{figure}
\includegraphics[width=\the\hsize]{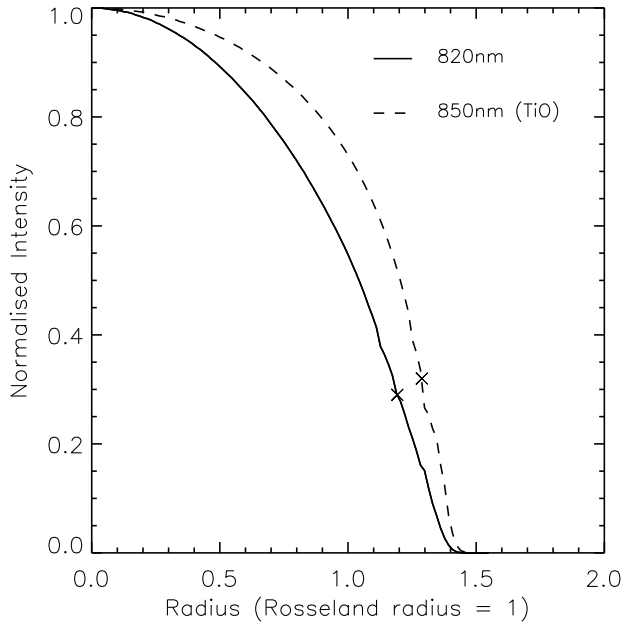}
\caption{Centre-to-limb variations (CLVs) for the 1M$_{\odot}$ static M giant model, X280, of BBSW. This shows the effect of both limb-darkening and molecular absorption in a TiO (850\,nm) band and a pseudo-continuum (820\,nm) band as a function of radius (normalised to the Rosseland radius). The $\tau_{\rm filter}=1$ radius for each passband is marked by a cross. }
\label{normalclvpair}
\end{figure}

Fig.~\ref{normalclvpair} shows monochromatic CLVs for the 1M$_{\odot}$ static model (X280) at two wavelengths, with normalised intensity plotted against radius. Although the intensity scale is normalised to 1.0 the ratio of the central intensities at pseudo-continuum and TiO bands can be very high. In this case it is about 2.5, but in some models at some neighbouring TiO/pseudo-continuum bands it can reach 1000. The radius scale is normalised to the Rosseland radius for consistency with HS98 and HSW98. This normalisation is independent of wavelength. Fig.~\ref{normalclvpair} shows a pseudo-continuum CLV at 820\,nm and a neighbouring TiO band CLV at 850\,nm. The 820\,nm CLV is more strongly limb-darkened than the 850\,nm one. In each case the cross marks the $\tau_{\rm filter}=1$ point which, like the $\tau_{\rm \lambda}=1$ point, is not directly observable.

Based on this plot we can conclude, from both the overall CLV width and the $\tau_{\rm \lambda}=1$ points, that the star is larger in the TiO band than in the pseudo-continuum band.

\begin{figure}
\includegraphics[width=\the\hsize]{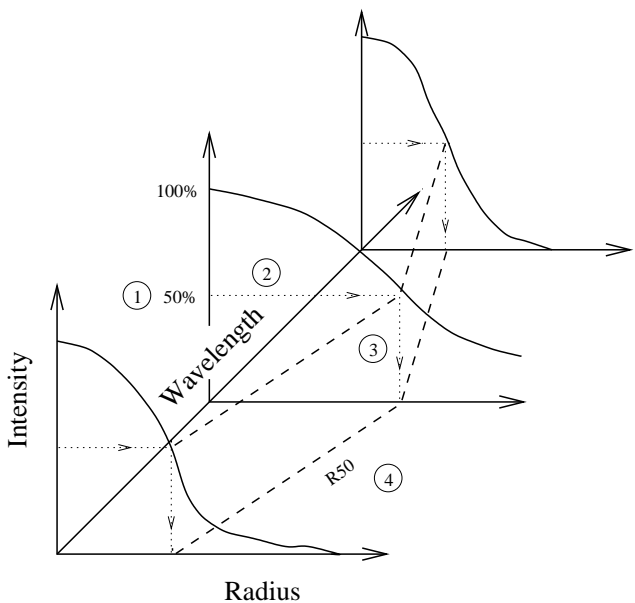}
\caption{The construction of the CLV ``spectrum''. See text.}
\label{clvspectraexample}
\end{figure}

\begin{figure*}
\includegraphics[width=\the\hsize, totalheight=20cm]{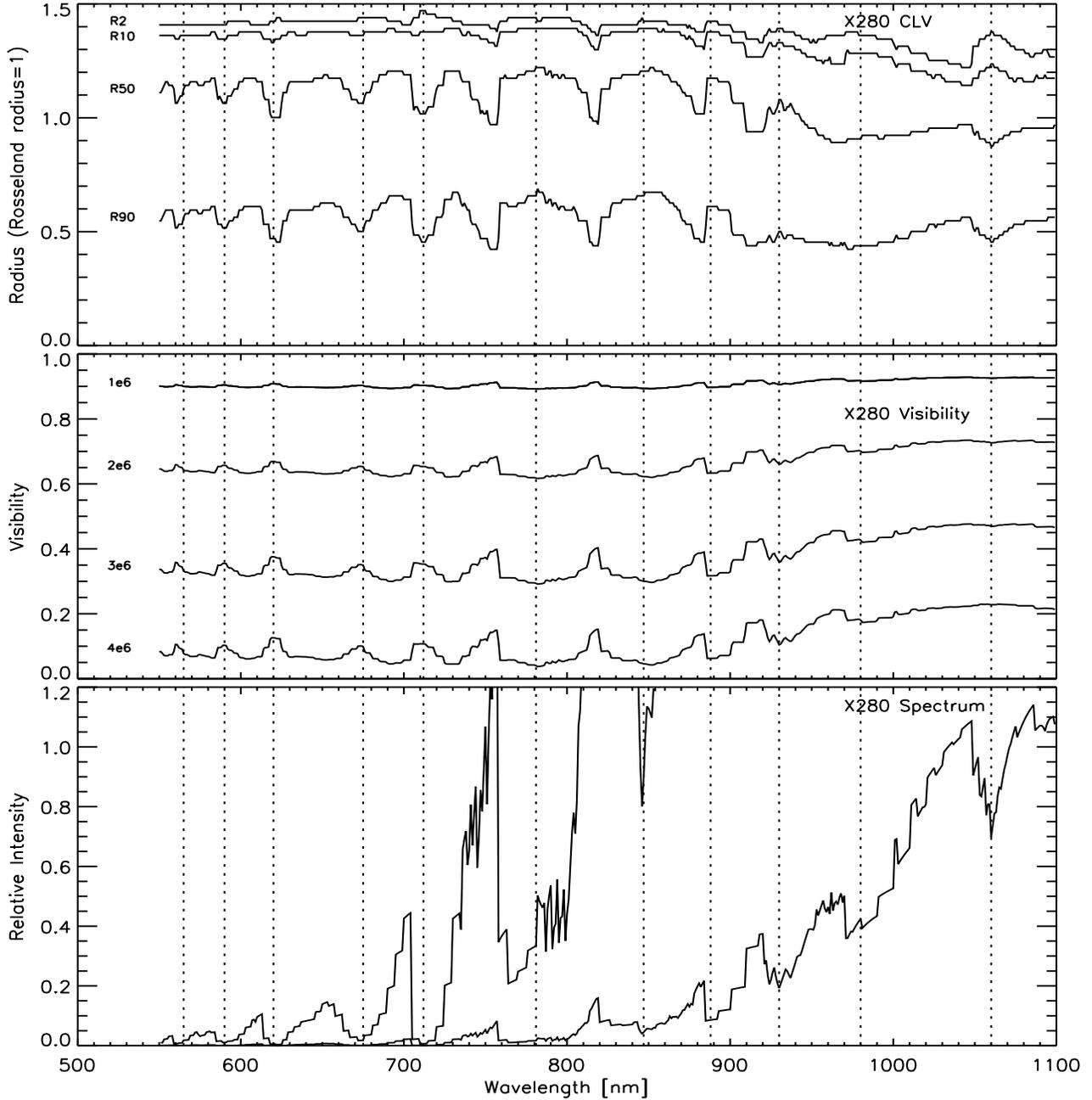}
\caption{ \emph{Top}: CLV spectrum for the 1M$_{\odot}$ static model, X280. This shows the CLV radius at 2, 10, 50 and 90\,\% of the central intensity, labelled R2, R10, R50 and R90 respectively. Each curve spans 550 monochromatic CLVs. The radius scale is normalised to the $\tau_{\rm Ross}=1$ radius. \emph{Middle}: Visibility spectra for the model at spatial frequencies of 1, 2, 3 and 4$\times 10^6$ rad$^{-1}$. The Rosseland radius is now assumed to be 25\,mas to fix the spatial frequency scale. \emph{Bottom}: Intensity spectrum for the model. The spectrum is plotted twice, with vertical scales differing by a factor of 20, to accommodate the large range in intensities. To guide the eye, dashed lines indicate the strongest absorption bands. }
\label{cvs}
\end{figure*}

In order to show wavelength-dependent CLVs most effectively, we have produced CLV ``spectra'' (i.e. plots of CLV versus wavelength). To understand these plots imagine placing many monochromatic CLVs (such as those in Fig.~\ref{normalclvpair}) beside each other as in Fig.~\ref{clvspectraexample}. We then choose a fraction of the the maximum (i.e. central) intensity, say 50\%, (point 1 on Fig.~\ref{clvspectraexample}) and find the points on the CLVs corresponding to this (point 2). We then project the resulting points onto the radius-wavelength plane (point 3) and draw a curve through them (point 4). For the case of the 1M$_{\odot}$ static model (X280) this produces the line labelled R50 in the top panel of Fig.~\ref{cvs}. Similarly we can form CLV spectra at 2, 10 and 90\,\% of the maximum intensity (labelled R2, R10 and R90 in Fig.~\ref{cvs}). This type of plot shows the overall shape and any low-level extension of each monochromatic CLV, and also the variation in these features with wavelength. The conventional intensity spectrum is included in the bottom panel of Fig.~\ref{cvs} to indicate the position of the various absorption bands. The middle panel is discussed in Section~\ref{modelvis}.

The shape of a CLV, such as that in Fig.~\ref{normalclvpair}, shows how the intensity varies as an observer's line-of-sight shifts from the disk's centre (where the observer sees deeper layers) towards the limb (higher layers). This variation includes the effects of both limb-darkening and molecular absorption. Therefore the R90 and R50 curves in Fig.~\ref{cvs} describe the compactness of the CLV. For this slightly-extended model atmosphere we see that, above about 750\,nm, the brightness is more concentrated towards the disk's centre at pseudo-continuum wavelengths than in TiO bands. The opposite is true below about 750\,nm. Only at very low intensity levels, say at R2, does one obtain meaningful intensity radii (cf. \citebare{BaSW91}), i.e. intensity radii that are physically related to the typical formation depths of observed spectral features. These show that the star is slightly larger in the light of TiO photons, that originate in its outer layers, than in the light of continuum photons that are formed deeper in its atmosphere. The width of the central portions of the CLV in either molecular bands or pseudo-continuum wavelengths depends on temperatures and temperature gradients and on the resulting intensity contribution functions in the relevant layers (S97). The assumption that an M giant star always appears ``larger'' when observed in a molecular band is too simple. This is discussed further in Section~\ref{anomolous}.

\subsection{Model-Derived Visibilities}
\label{modelvis}
An observer does not directly see the surface brightness distribution of a star, but can measure fringe visibility with an interferometer. If we could measure both the amplitude and phase of the visibility, and deconvolve the instrumental profile, we could reconstruct the surface brightness distribution from the sampled interferometric data. The phase information tells us about asymmetries in the brightness distribution but the only models available are radially symmetric and do not include the effects of, for example, hotspots. Therefore we will only consider the amplitude of the visibility in this paper.

We converted each monochromatic CLV to a monochromatic visibility curve using the Hankel transform, which gives the response of a one-dimensional interferometer to a circularly symmetric disk (see \citebare{H97}). To fix the spatial frequency scale we must assume an angular size for the model. Following BBSW and BSW96 we have used the $\tau_{\rm Ross}=1$ and the parent-star $\tau_{\rm Ross}=1$ points (for static and Mira models respectively) to define the radius. We assumed this radius to be 25\,mas. This is the approximate size of the stars referred to in this paper. Note that the Rosseland radius of a Mira model at a given phase is often close, but not identical, to the Rosseland radius of the parent-star model. 

\begin{figure}
\includegraphics[width=\the\hsize]{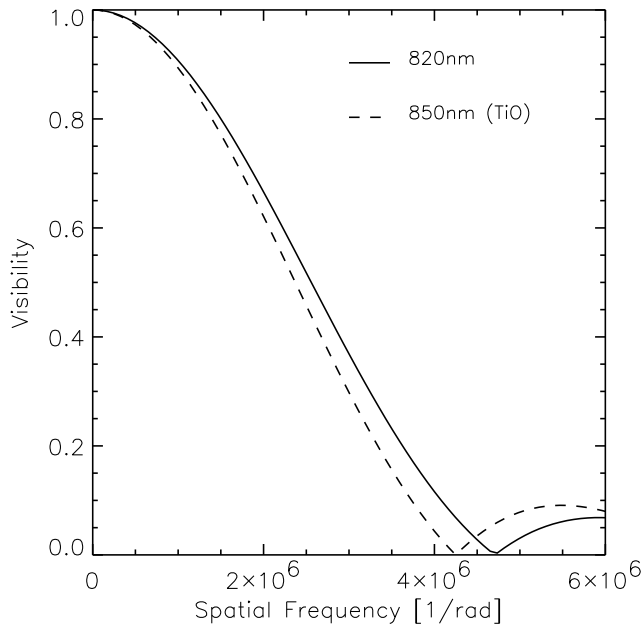}
\caption{Two visibility curves for the 1M$_{\odot}$ static M giant model, X280, of BBSW. These are Hankel transforms of the CLVs in Fig.~\ref{normalclvpair}. To fix the spatial frequency scale we have assumed that the star has an angular diameter of 50\,mas at the Rosseland radius point in Fig.~\ref{normalclvpair}. }
\label{normalvispair}
\end{figure}

Fig.~\ref{normalvispair} shows the visibility curves for the two monochromatic CLVs of Fig.~\ref{normalclvpair}, with visibility plotted against spatial frequency. Spatial frequencies are measured in rad$^{-1}$ and they are numerically equivalent, for a 2-element interferometer, to the baseline length measured in wavelengths. Based on this plot we see that the star is larger in the TiO band because its visibilities are lower at all spatial frequencies out to the first null. This concurs with our conclusion from Fig.~\ref{normalclvpair}. If we had the complete visibility curves we could also reconstruct the true shape of the CLVs.

We next used the monochromatic visibility curves to produce a visibility ``spectrum'' (i.e. visibility versus wavelength) at several different spatial frequencies just as we made the CLV spectrum. This is shown in the middle panel of Fig.~\ref{cvs}. These curves look somewhat like inverted versions of the CLV spectra.

\subsection{Broad-band Visibilities}
\label{smoothing}
In order to compare the predictions of the models with observations we must include the effect of a real, broad-band, filter on the shape of the CLV. This is usually done by including the filter profile in the calculation of each narrow-band CLV. However, to make wavelength-dependent CLVs for many different filter profiles (or passbands) in this way and subsequently make a CLV or visibility spectrum would be time consuming. Instead, we produced the monochromatic CLV and visibility spectra, such as in Fig.~\ref{cvs}, and then simply smoothed them by convolving each with the required observing passband and then normalising by the intensity spectrum, i.e.,
\begin{equation}
S_{conv}=\frac{(S \cdot I) \otimes f}{ I \otimes f}
\label{convoleqn}
\end{equation}
where $S_{conv}$ is the smoothed CLV or visibility spectrum, $S$ is the monochromatic CLV or visibility spectrum, $I$ is the intensity spectrum, $f$ is the observing passband function and $\otimes$ denotes convolution. This was done individually for each radial position (i.e. for each R$\it x$ in Fig.~\ref{cvs}) and for each spatial frequency.

\begin{figure*}
\includegraphics[width=\the\hsize, totalheight=20cm]{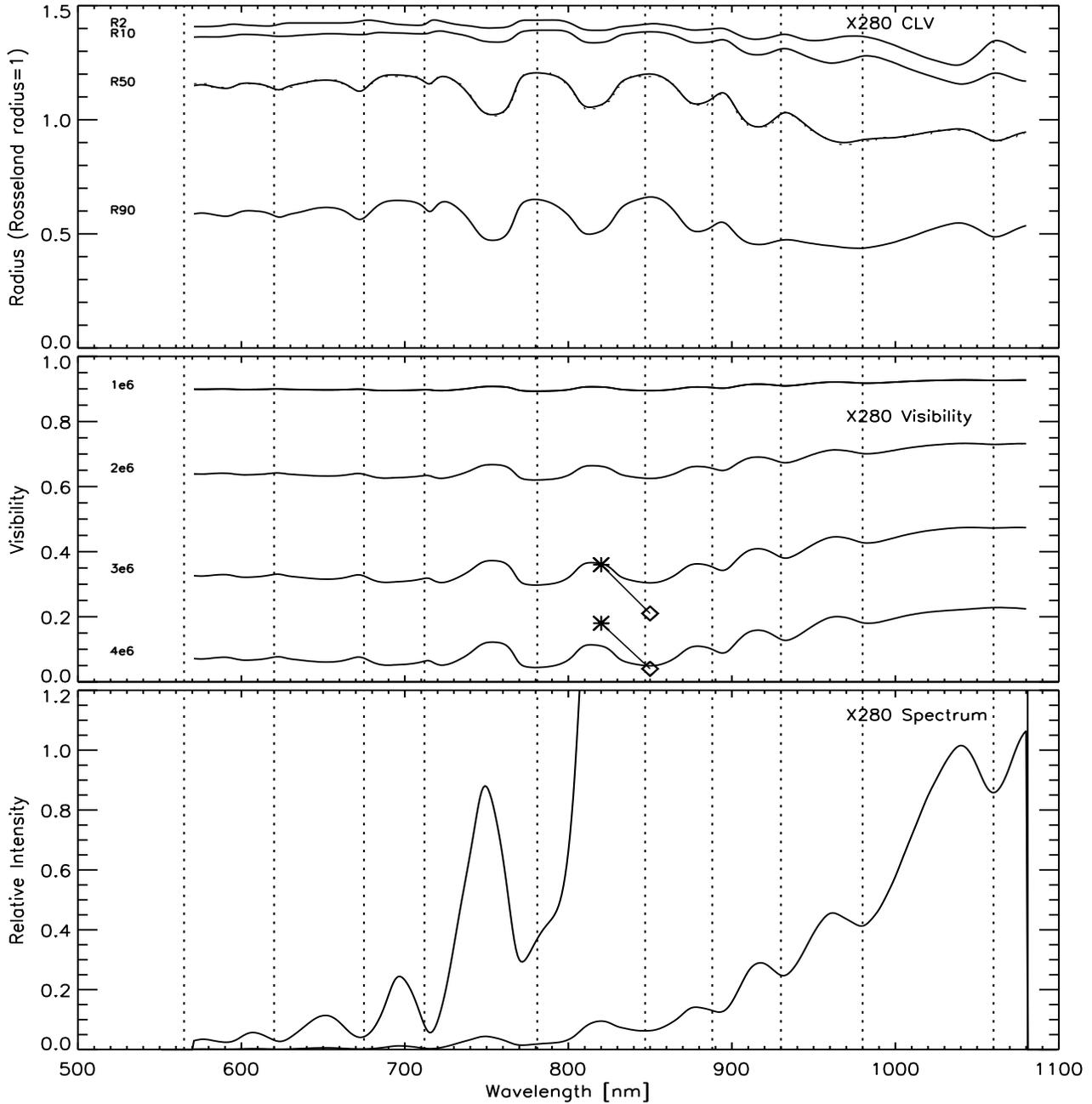}
\caption{As for Fig.~\ref{cvs} but smoothed to simulate the effect of observing with a 20\,nm FWHM super-Gaussian passband. The dotted curve overlying the R50 CLV spectrum is discussed in Section~\ref{smoothing}. Superimposed on the visibility spectra are data for R~Dor (asterisks and diamonds) interpolated from Fig. 3 of \protect\citeone{JBR97}. These data are discussed in Section~\ref{opvs}. }
\label{cvsconv}
\end{figure*}

As an example we present Fig.~\ref{cvsconv}. Here the 1M$_{\odot}$ static (X280) model's CLV and visibility spectra from Fig.~\ref{cvs} have been smoothed, using Eqn.~\ref{convoleqn}, to simulate the effect of a super-Gaussian (i.e. $e^{-x^4}$) filter with a 20\,nm FWHM. The intensity spectrum was smoothed by a simple convolution with the same filter profile. This profile approximates quite well the interference filters that are often used. The superimposed points from \citeone{JBR97} are discussed in Section~\ref{opvs}.

To check the accuracy of this method we compared broad-band CLV and visibility spectra for the X280 model, that were produced by smoothing with Eqn.~\ref{convoleqn}, to spectra produced directly by including the filter profile in the calculation of each narrow-band CLV. The difference between the CLV spectra produced by these methods was generally below 2\% and the difference between the visibility spectra was below 0.5\%. In Fig.~\ref{cvsconv} a CLV spectrum at 50\% of the central intensity, that was produced by the direct method, is plotted. It is almost indistinguishable from the R50 curve. A visibility spectrum, produced by the direct method, is plotted at 2$\times 10^6$ rad$^{-1}$ and is indistinguishable from the spectrum smoothed using Eqn.~\ref{convoleqn}. A check using the E8380 model produced similar results.


In going to a smoothed plot, Fig.~\ref{cvsconv} shows that there remain obvious variations in the visibility which should be observable. Also, multi-wavelength visibilities as measured by MAPPIT provide many more data points for comparing models with observations than do present narrow-band techniques. Clearly this provides a test of how well the models predict the variation in visibility with wavelength at a single epoch. The narrow-band work done to date, by us and other authors, has only been a test of how well the models predict the variation of visibility with spatial frequency in that band. Although several bands are normally used in narrow-band work they are sometimes observed at slightly different epochs. In the intervening time the phase, and hence the visibility, may change (see e.g. the ``D'' models in Fig.~\ref{RCasWBH96}).

\section{The Observational Perspective}
\label{obsperspective}
An interferometer measures the fringe visibility, but only to a limited spatial frequency. This limited range, combined with measurement errors and the practical difficulties involved in detecting high-spatial-frequency low-contrast fringes, makes it difficult to directly recreate the true CLV of a star. Instead observers rely on model fitting. Standard limb-darkening profiles (such as UD, FDD or Gaussian) or CLVs derived from model photospheres, such as those of BBSW, are converted to visibilities and fitted to the observed visibility curves. Only at this point can we determine a value for the stellar ``radius''. This could be the UD or FDD radius, some reasonable multiple of the HWHM of the Gaussian, the $\tau_{\lambda} = 1$, $\tau_{\rm filter}=1$ or $\tau_{\rm Ross}=1$ optical-depth radius, or an intensity radius deduced from the shape of the CLV.

There are several problems that arise in this process. Firstly, there are known modelling limitations. Observations have shown that M giant and supergiant disks exhibit bright surface features and/or geometric elongations (see e.g. \citebare{WBB92}), but no model yet incorporates these phenomena. The BBSW models also have limited reliability in deep absorption bands.

Secondly, most of the BBSW and BSW96 model CLVs show a null in their visibilities at spatial frequencies accessible to 4\,m-class telescopes but most observations do not show it (see e.g. \citebare{HST95}, \citebare{WBH96}, \citebare{BZvdL97}). Instead, Gaussian curves provide better fits to the observations. \citeone{HST95} suggested that this could be due to localized hotspots or circumstellar scattering that increases the visibility. If this is the case then comparing multi-wavelength visibilities with model predictions, at low spatial frequencies which are relatively unaffected by these phenomena, could provide a way to improve the models until these effects are included.

\begin{figure*}
\includegraphics[width=\the\hsize, totalheight=20cm]{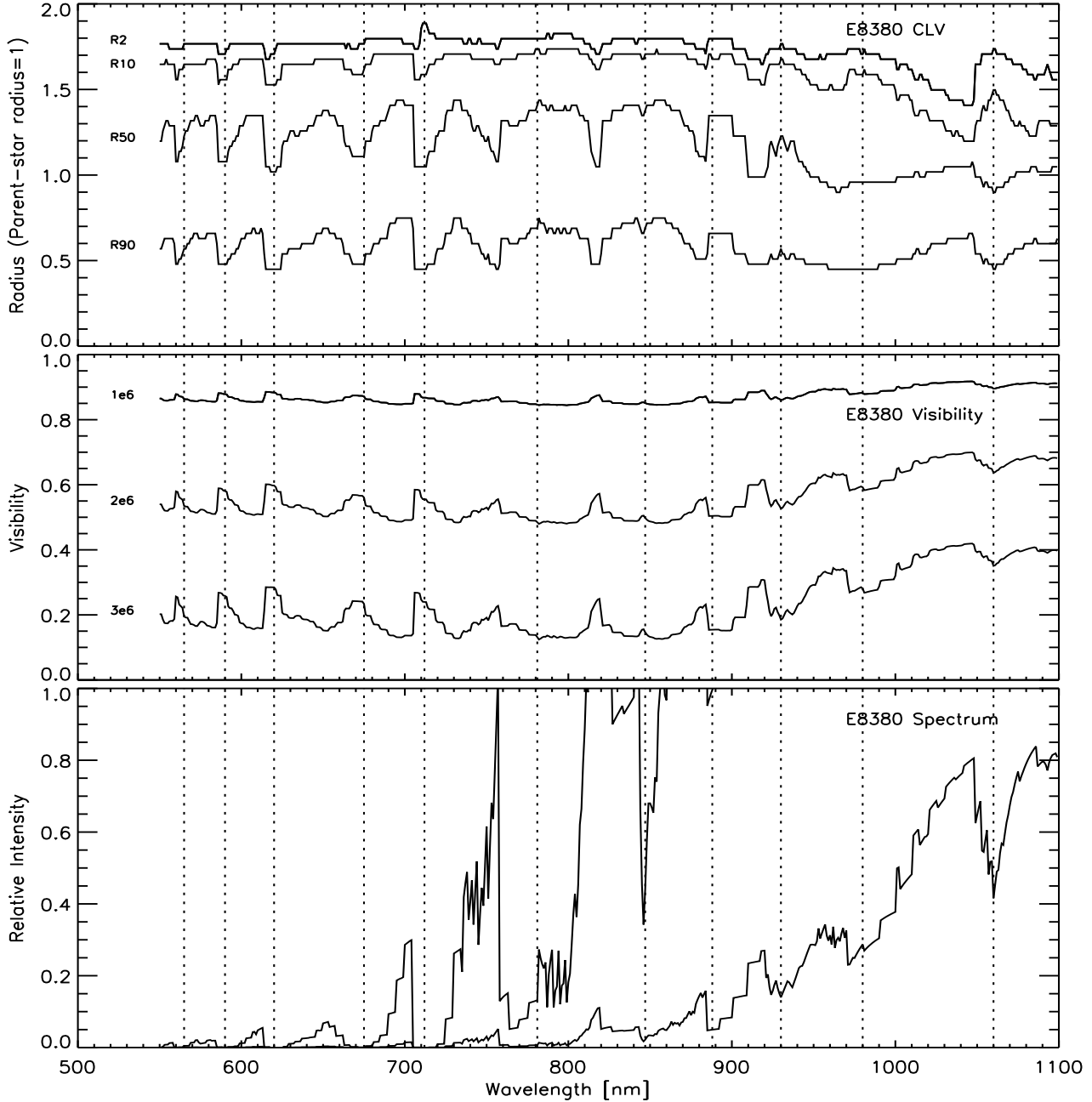}
\caption{As for Fig.~\ref{cvs} but for the 1M$_{\odot}$ Mira model, E8380. Here the radius is normalised to the parent star's Rosseland radius, $R_p$. }
\label{cvsE8380}
\end{figure*}

Thirdly, an observer's interpretation of the data is hindered by the similarity of the visibility curves within the first null even when the CLVs are significantly different. For example, in Fig.~\ref{cvs} the visibilities between 1000 and 1100\,nm, which includes a VO band, are approximately constant with wavelength, even though the CLV shape varies significantly across this band. In this case it is the limb-darkening that differs. Fig.~\ref{cvsE8380} shows the CLV, visibility and intensity spectra for a 1M$_{\odot}$ Mira model. Here the visibility spectrum has a dip at the VO band. Since the models are expected to give reasonable predictions at these wavelengths (\citebare{B90}), this part of the visibility spectrum could be useful in distinguishing between different models, despite the difficulty in observing these wavelengths. Additionally, the similarity in the visibility curves means that if the diameter is treated as a free parameter it can be impossible to tell whether two visibility curves are different due to different diameters (given a single limb-darkening profile) or to different limb-darkening profiles themselves, unless one looks beyond the first null.

Finally there is the possible misinterpretation of model visibilities by an observer. A low visibility within the first null, at a particular spatial frequency, means the interferometer is resolving the source more than if the visibility was high. We interpret this to mean the source is larger in angular size. If we assume the TiO overlies the pseudo-continuum forming layers, as it must do to form absorption bands, then we would expect to see a larger star in the TiO bands than in the pseudo-continuum bands. Observationally we would expect lower visibilities (within the first null) in TiO bands. However, in some  models in some TiO bands we noticed that the visibility was higher than in neighbouring pseudo-continuum bands. See, for example, the visibilities below about 750\,nm in Fig.~\ref{cvs} and Fig.~\ref{cvsE8380}. A first glance at the model visibility curves might lead an observer to conclude that the star is smaller in these TiO bands than in the neighbouring pseudo-continuum bands. This does not seem to make sense. The explanation lies in the form and extent of the limb-darkening present (see section~\ref{anomolous}). The observer's misinterpretation of these model visibilities results from not considering carefully the visibilities near and beyond the first null or at high spatial frequencies.

Clearly the visibility spectrum, and monochromatic visibility curves, require careful interpretation.

\section{An Extreme Limb-Darkening Effect in Deep Titanium-Oxide Bands}
\label{anomolous}
A star with a $\it compact$ atmosphere, such as the Sun, has a relatively sharp edge to its brightness distribution. The distance of this edge from the disk's centre is effectively independent of wavelength and thus defines the stellar radius. If observations of a $\it compact$-atmosphere star appear to imply wavelength-dependent radii this indicates incorrect assumptions have been made about limb-darkening. In fact, this provides a test for whether or not we have chosen the correct limb-darkening form at each wavelength. For example, if a single limb-darkening model (e.g. UD) is fitted to observed visibilities at two wavelengths that actually have different CLVs this will, incorrectly, give different angular diameters.

In the case of an extended-atmosphere star this effect may occur in conjunction with a genuine physical difference in the extension of the stellar disk seen at different wavelengths. The CLV in a strong-absorption spectral feature formed in high atmospheric layers (i.e. large $\tau_{\lambda}=1$ radius) may be flatter than the CLV in a pseudo-continuum feature formed in deep layers (small $\tau_{\lambda}=1$ radius). For a small atmospheric extension, such as occurs in most static M giants, the CLV behaviour (e.g. \citebare{WK79}, ST87, HS98) may obscure the larger size of the star at the strong-absorption wavelength. Nevertheless, the geometric extension effect normally prevails in very cool non-Miras and in Miras having substantially extended atmospheres. In this case the central maximum of the visibility is narrower at the wavelength of the strong-absorption feature which indicates a larger disk. Data reduction by means of an identical limb-darkening model would lead to an underestimate of the ratio of the disk sizes in strong and weak absorption, but this would not suggest a smaller disk at the stronger spectral absorption feature.

We found, however, during the present investigation that some of the BSW96 and HSW98 Mira models predict extreme CLV shapes in very strong TiO bands. We use the term ``extreme'' here for the ``narrow-bright-core'' effect described in HSW98. To save space not all of these models  are presented here, although the 1M$_{\odot}$ Mira model (E8380) in Fig.~\ref{cvsE8380} is a typical example. The width of the central maximum of the visibility in very strong TiO bands (e.g. 710\,nm) becomes significantly broader than that seen at neighbouring pseudo-continuum wavelengths, a situation that at first glance seems incorrect. The disk, if seen directly by an observer, would be extremely limb-darkened and have very extended faint outer portions. The CLV has a Gaussian shape and the $\tau_{\lambda}=1$ optical-depth radius is positioned in the low-intensity wing (see e.g. Fig.~8 of HSW98). Only very high accuracy measurement of the visibility curve at high spatial frequencies would reveal the large size of the stellar disk in the TiO band.

\begin{figure*}
\includegraphics[width=\the\hsize, totalheight=20cm]{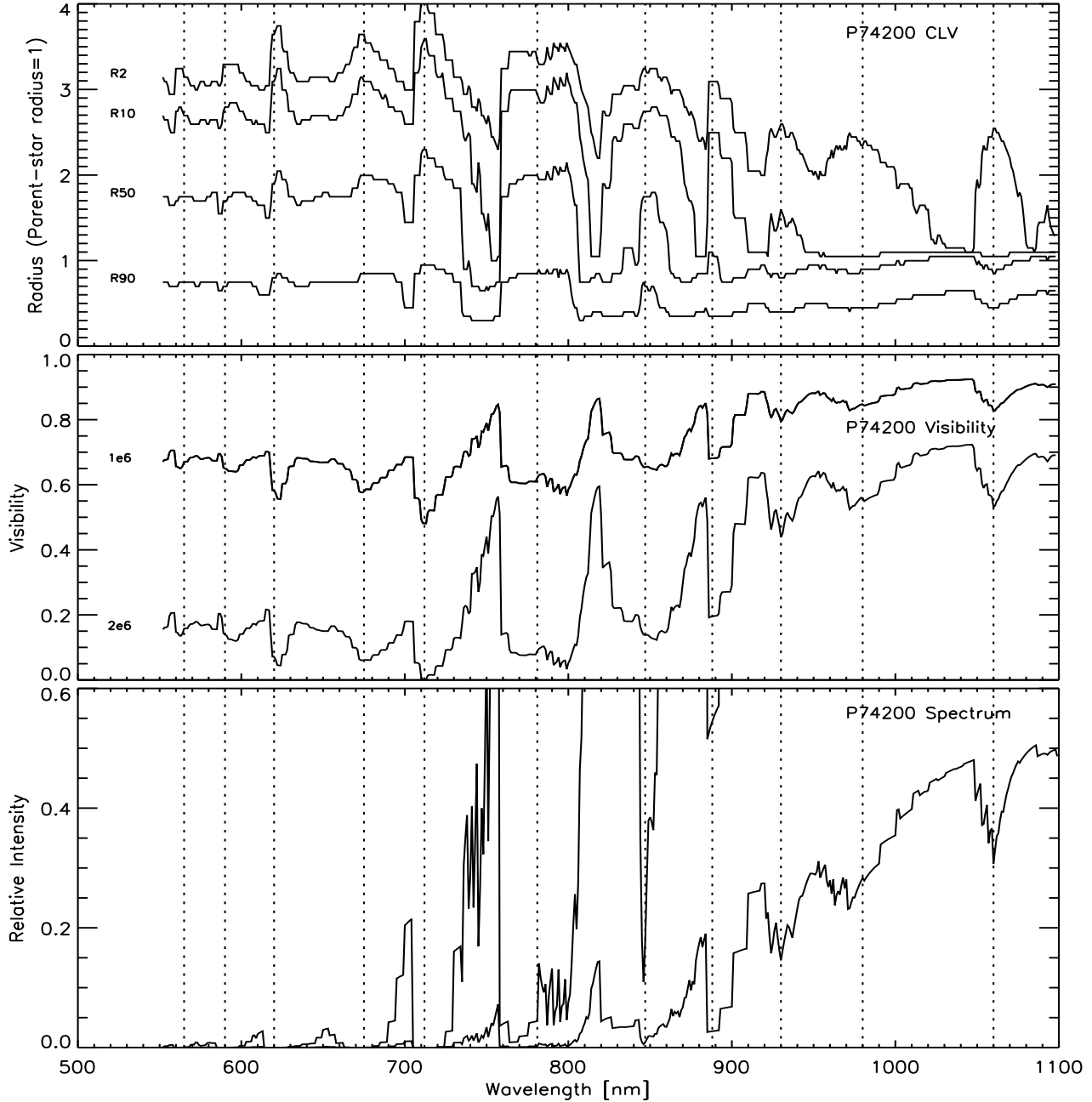}
\caption{As for Fig.~\ref{cvs} but for the 1M$_{\odot}$ Mira model, P74200. Here the radius is normalised to the parent star's Rosseland radius, $R_p$. The extreme limb-darkening effect is not present but molecular absorption has caused large differences in the star's size between absorption and pseudo-continuum bands. }
\label{cvsP74200}
\end{figure*}

This extreme limb-darkening effect may also show up in some very cool non-Mira models (see Fig.~\ref{cvs}). However, it is most pronounced in certain Mira models such as the near-maximum 1M$_{\odot}$ model (E8380). Fig.~\ref{cvsE8380} shows the extreme central concentration of the strong-TiO band brightness distributions that occur below about 750\,nm in this model. Other extreme examples are found at near-minimum phases of the P series of models of HSW98 (see the P71800 model in Fig.\,2, upper left panel, of S97). However, the near-maximum model (P74200) of the same series (Fig.\,2, upper right panel, of S97) does \emph{not} show this effect. Fig.~\ref{cvsP74200} shows, for this near-maximum model, the ``expected'' situation, i.e. the central visibility maxima in strong TiO bands are narrower than those in neighbouring pseudo-continuum bands. Note, all Mira model CLVs in this paper are displayed with their radii normalised to their parent star's Rosseland radius.

Unfortunately, this effect depends not only on the specific model and model phase but also on subtle details of the structure of the upper layers (HSW98). Present Mira models are still approximate. The temperature stratification in the upper atmosphere is significantly affected by the simple treatment of TiO molecular band absorption which is particularly sensitive for strongly saturated bands (see BBSW, BSW96). This simple treatment also does not account for the effects of Doppler-shifting in layers with outflowing and infalling matter. Furthermore, the adopted absorption coefficients  directly affect the CLV shape.

A search of the literature revealed only two sets of published visibility curves relevant to this discussion. Fig. 2b of \citeone{HST95} shows visibilities for R~Cas at 700\,nm and 710\,nm obtained in September 1993. Fig. 4 of \citeone{WBH96} also shows visibilities for R~Cas at 700\,nm and 714\,nm obtained in September 1994. There is no evidence for an extreme limb-darkening effect in this star. We have MAPPIT data for R~Dor covering these two bands which we are currently processing. Therefore, it is not clear at present whether this extreme limb-darkening effect does occur in real Mira stars or is only a modelling artifact produced by inadequate modelling techniques.

\begin{figure*}
\includegraphics[width=\the\hsize]{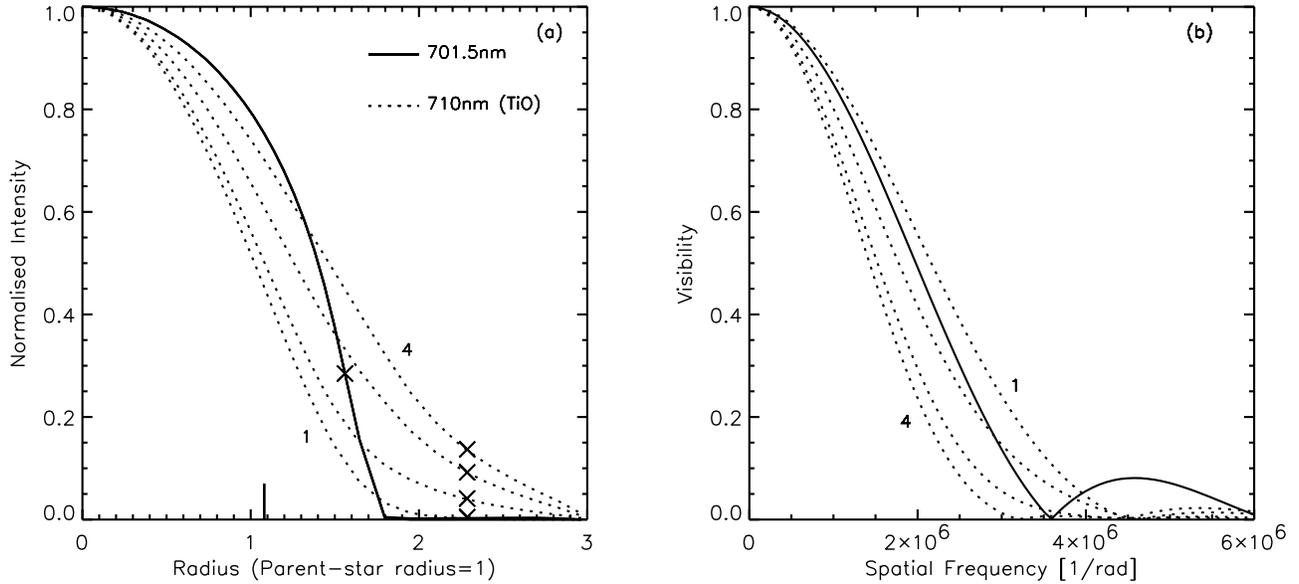}
\caption{CLVs (a) and visibilities (b) for the 1M$_{\odot}$ Mira model, E8380, at 701.5 (pseudo-continuum) and 710\,nm (TiO) showing how the TiO band CLV and visibility change as the temperature gradient in the outer layers is artificially flattened (see text). Crosses mark the $\tau_{\lambda}=1$ radii points on the CLVs and the $\tau_{1.04}=1$ continuum radius is marked by a short dash at $1.08\times R_p$. }
\label{many-clv}
\end{figure*}

A demonstration of the influence of the temperature gradient on the brightness profile is given in Fig.~\ref{many-clv}. We calculated CLVs for the Mira model (E8380) in two rectangular passbands of 1\,nm width. One was centered at 701.5\,nm in a pseudo-continuum band and the other at 710\,nm in a very strong TiO band. Fig.~\ref{many-clv}a shows that the very strong TiO CLV (dashed curve 1) is much narrower than the pseudo-continuum CLV (solid curve). In Fig.~\ref{many-clv}b the visibility at the very strong TiO band (dashed curve 1) is broader than that at the pseudo-continuum band (solid curve). This is what we would expect after seeing Fig.~\ref{many-clv}a, yet an observer, who only sees the visibility curves and usually only within the first null, might naively conclude that the star was larger in the pseudo-continuum band.

The respective $\tau_{\lambda}=1$ radii are 2.29 and 1.56 $\times R_{\rm p}$ (marked by crosses) with the former lying in the very dim outer portions of the CLV curve. So, by the $\tau_{\lambda}=1$ definition, the TiO band radius $\it is$ larger than the pseudo-continuum radius. Note that the $\tau_{1.04}=1$ continuum radius is at 1.08 $\times R_{\rm p}$ (HSW98).

We now perform a simple numerical experiment to investigate the effect on the brightness profile of changing the temperature gradient in the outer layers. We assume that the outermost atmosphere is optically thin with respect to features affecting radiative equilibrium above 2.0, 1.9 and 1.85 $\times R_{\rm p}$ (where $T$ = 1090, 1180 and 1250\,K) sequentially. This has the effect that the temperature gradient becomes flatter and  follows a $T(r) \propto r^{-1/2}$ law. We may consider this as, for example, a very simplistic correction for possibly excessive near-surface cooling which could be caused by over-blanketing due to inadequate molecular band treatment. The result is shown in Fig.~\ref{many-clv}. The 701.5\,nm pseudo-continuum CLV (FDD-like shape) is not affected by these manipulations. On the other hand, the width of the very-strong-TiO CLV (Gaussian shape) becomes progressively broader with a corresponding change in the visibility curve (dashed curves 2 to 4). Flattening the gradient above any layer that lies inside about 1.90 $\times R_{\rm p}$ (dashed curve 2) results in the ``expected'' situation where the strong-TiO visibility is narrower than the pseudo-continuum visibility.

It is obvious from the above that accurate interferometric observations of strong TiO bands would provide a sensitive method of probing the temperature structure of the outermost atmospheric layers.

\section{Comparison with Published Visibilities}
\label{comparison}
\subsection{The 850\,nm TiO Band in R~Dor}
\label{obs-820/850}
\citeone{BZvdL97} measured visibility curves for R~Dor using an interference filter of 40\,nm width centered at 855\,nm. The pulsation phase of the star at that time is difficult to determine due to the recent irregularity of R~Dor (see \citebare{BZJ98}). \citeone{JBR97} found very similar visibilities to the \citeone{BZvdL97} results but at a pulsation phase of about 0.7 and with a bandwidth of 20\,nm centred at 850\,nm. These visibilities agreed very well with each other despite the different observing passbands. Both the static (X) and Mira (E) models used in this paper to represent R~Dor can be used to explain this agreement. Fig.~\ref{cvs} and Fig.~\ref{cvsE8380} show that the visibility spectra at all spatial frequencies are approximately constant from 820\,nm to 870\,nm. This is also the case for other phases of the E model. Thus, as long as the wings of the filter are within this region, the measured visibility will be independent of the filter width and central wavelength.

In fact the visibility spectrum could be a useful tool for choosing the filter width, its central wavelength and the allowable leakage from, for instance, an absorption band. Observers normally use the intensity spectrum to choose filter properties. 

\subsection{Other Multi-wavelength Visibilities}
\label{opvs}
We are aware of only four sets of published multi-wavelength visibilities that we can compare with the BBSW models using our visibility spectra. They are for R~Dor \cite{JBR97}, R~Cas (\citebare{HST95} and \citebare{WBH96}) and R~Leo \cite{THB94}.

\begin{figure*}
\centerline{\includegraphics[bb=75 535 509 700, clip=true]{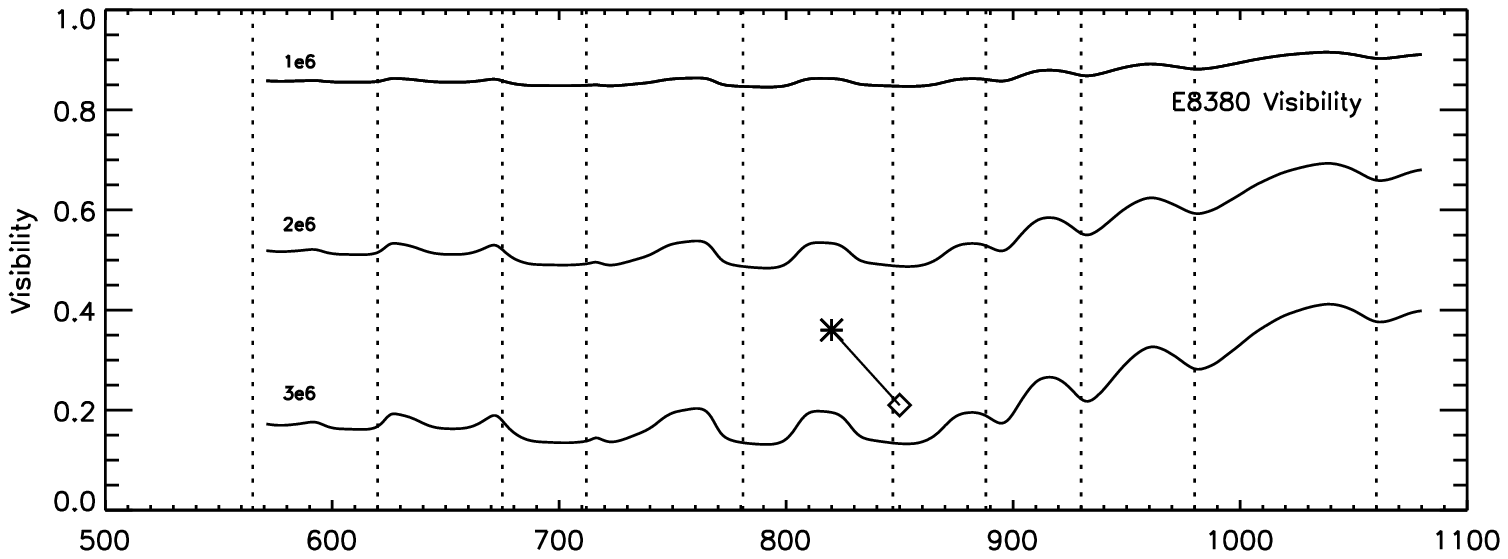}}
\caption{Observations of R~Dor by \protect\citeone{JBR97}, compared with a solar-mass Mira model. }
\label{RDorE8380}
\end{figure*}

\begin{figure}
\centerline{\includegraphics[bb=75 546 300 694, clip=true]{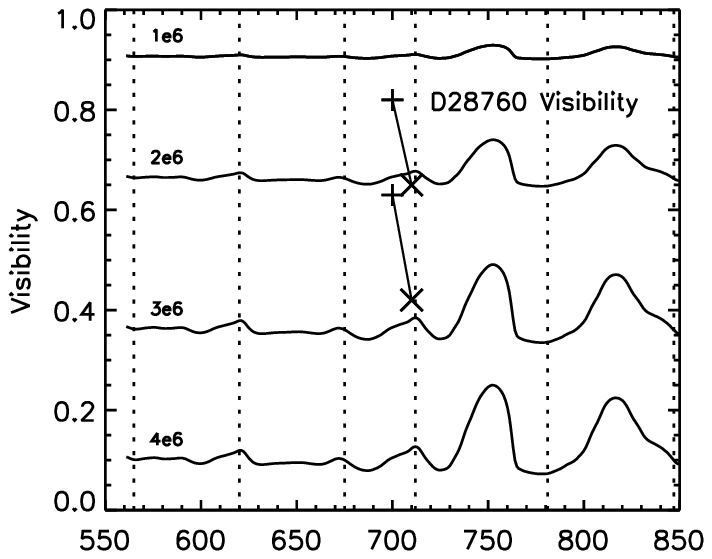}}
\centerline{\includegraphics[bb=75 535 300 694, clip=true]{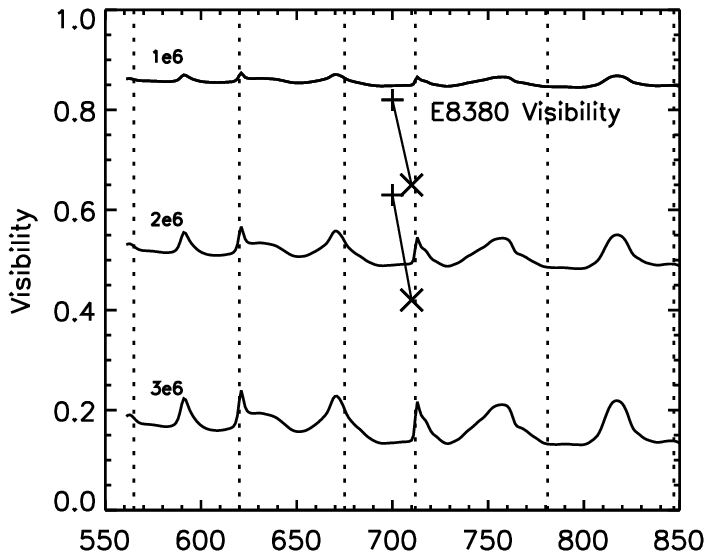}}
\caption{Observations of R~Cas by \protect\citeone{HST95}, compared with two solar-mass Mira models of different pulsation modes: fundamental (top) and first-overtone (bottom). }
\label{RCasHST95}
\end{figure}

\begin{figure}
\centerline{\includegraphics[bb=75 546 275 694,clip=true]{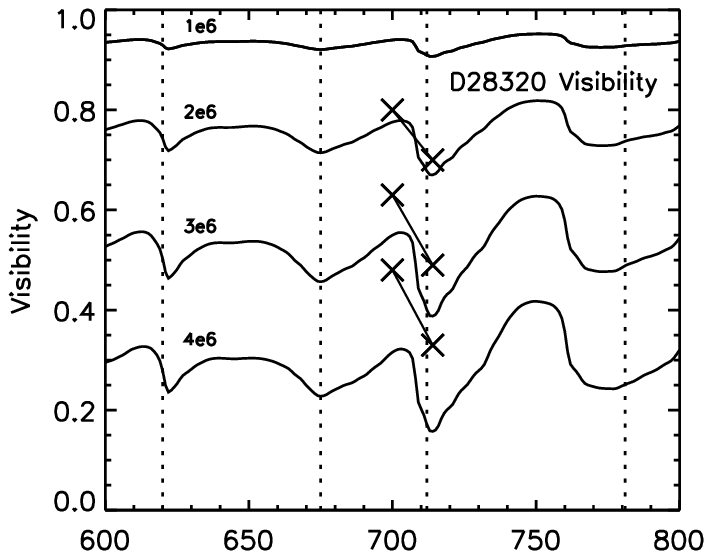}}
\centerline{\includegraphics[bb=75 546 275 694,clip=true]{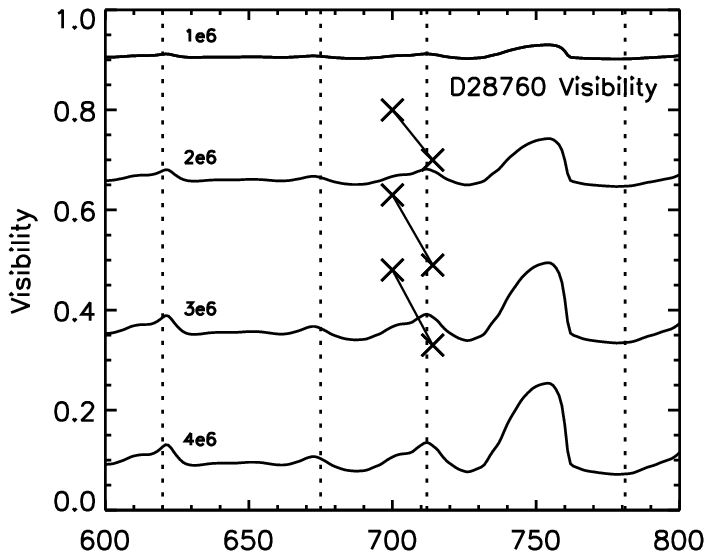}}
\centerline{\includegraphics[bb=75 535 275 694,clip=true]{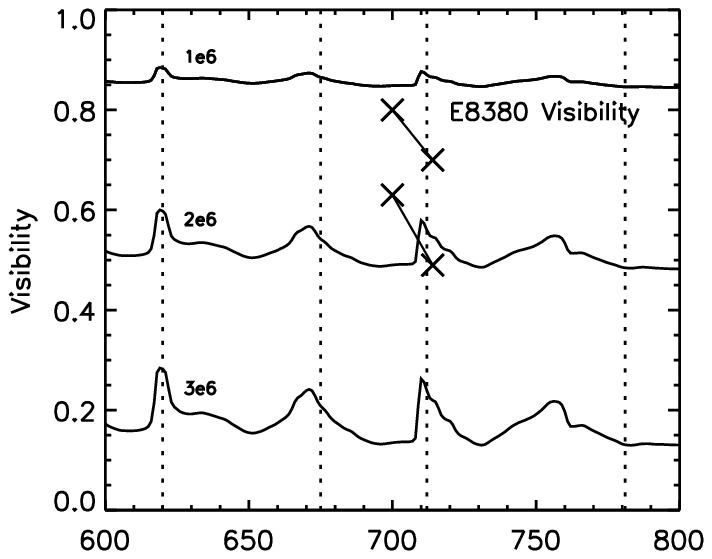}}
\caption{Observations of R~Cas by \protect\citeone{WBH96}, compared with three solar-mass Mira models of different pulsation modes: fundamental (top \& middle) at different phases and first-overtone (bottom). }
\label{RCasWBH96}
\end{figure}

\begin{figure}
\centerline{\includegraphics[bb=75 546 300 694,clip=true]{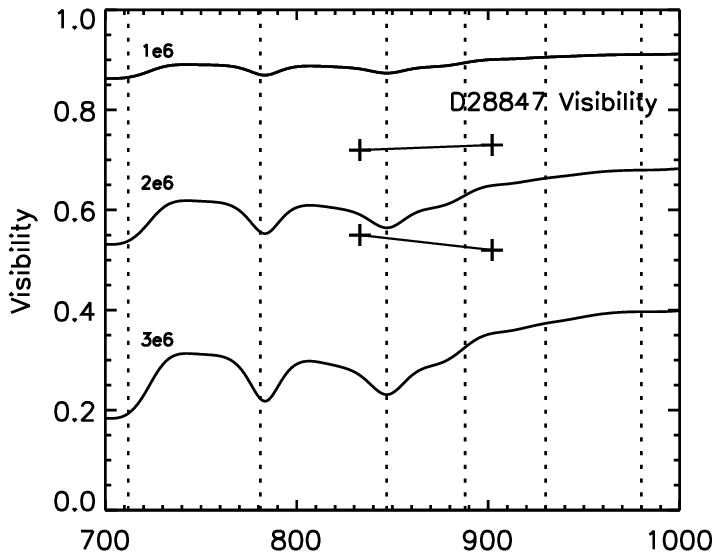}}
\centerline{\includegraphics[bb=75 535 300 694,clip=true]{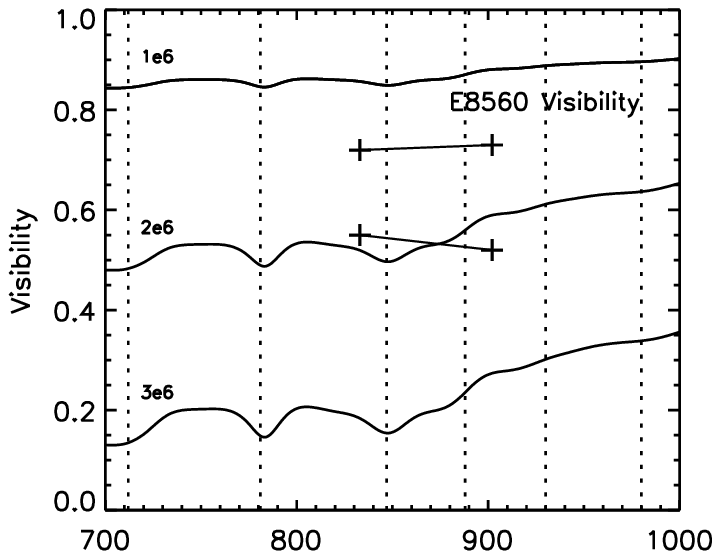}}
\caption{Observations of R~Leo by \protect\citeone{THB94}, compared with two solar-mass Mira models of different pulsation modes: fundamental (top) and first-overtone (bottom). }
\label{RLeo}
\end{figure}

Our criterion for choosing these sets of observations was that they include adjacent TiO/pseudo-continuum bands at the same epoch (i.e. same pulsation phase). Figs.~\ref{cvsconv} and \ref{RDorE8380} to~\ref{RLeo} show these observations over-plotted onto relevant models. In each plot, the visibility spectra have been smoothed, using Eqn.~\ref{convoleqn}, to closely match the FWHM of the filters used for the observations. The data points are linearly interpolated from the relevant figures in each paper where necessary. We have not included errors in these plots but the errors in the measured data ranged from 6 to 20\,\%. Straight lines between data points simply connect points at the same spatial frequency for clarity. The model visibilities are plotted for either a Rosseland (for static models) or parent-star Rosseland (for Mira models) radius of 25\,mas for simplicity, i.e. no attempt has been made to match the size of the model star to the size of the observed star. Therefore, this comparison shows to what degree the models predict the variation in visibility with wavelength, but not how well they predict the absolute visibility.

The \citeone{JBR97} data for R~Dor are plotted on the visibility spectra for the 1M$_{\odot}$ static model (X280) in Fig.\ref{cvsconv}. The asterisks are observations at 820/18\,nm and the diamonds are observations at 850/20\,nm. The upper and lower pairs of points are at spatial frequencies of $3$ and $4 \times 10^6$ rad$^{-1}$ respectively. The model has a much smaller variation in visibility with wavelength than the observations. Since R~Dor shows some Mira-like properties we compare the same data to the 1M$_{\odot}$ Mira model (E8380) of BBSW in Fig.\ref{RDorE8380}. Only data points at a spatial frequency of $3 \times 10^6$ rad$^{-1}$ have been plotted here and the model has been smoothed to simulate the effect of a 20\,nm FWHM super-Gaussian filter. Again the model has a much smaller variation in visibility with wavelength than the observations.

\citeone{HST95} observed the Mira variable R~Cas at 700/10 and 710/10\,nm in September 1993 at a phase of about 0.03. In Fig.~\ref{RCasHST95} we compare these observations to a fundamental-mode Mira model (D28760) at phase 0.0 (top) and to a first-overtone mode Mira model (E8380) at phase 0.0 (bottom). These models have a phase close to the observations and were used in \citeone{HST95} to represent R~Cas. The upper and lower pairs of points in each panel are at spatial frequencies of $2$ and $3 \times 10^6$ rad$^{-1}$ respectively. Again the observed visibilities vary significantly more than the model visibilities. In fact, the models predict the reverse of the observed variation.

\citeone{WBH96} observed R~Cas at 700/6 and 714/6\,nm in September 1994 at a phase of about 0.91. In Fig.~\ref{RCasWBH96} we compare these observations to fundamental-mode Mira models at phase 1.8 (D28320: top) and phase 0.0 (D28760: middle) and to a first-overtone mode Mira model (E8380) at phase 0.0 (bottom). Note that both Mira models and Miras themselves show cycle to cycle variations in their behaviour hence pulsation phases can be greater than 1.0 (BSW96, HSW98). For the two fundamental-mode models the top, middle and bottom pairs of points in each panel are at spatial frequencies of $2$, $3$ and $4 \times 10^6$ rad$^{-1}$ respectively. For the first-overtone model only the observations at spatial frequencies of $2$ and $3 \times 10^6$ rad$^{-1}$ are shown. Only the fundamental-mode model approaching maximum brightness in its second cycle (i.e. at phase 1.8) has a similar variation in visibility with wavelength to the observations. The fundamental model at maximum brightness (i.e. phase 0.0) and the first overtone model are not well matched to the observations and again show the reversed behaviour.

\citeone{THB94} observed the Mira variable R~Leo at 833/41 and 902/50\,nm in January 1992 at a phase of about 0.26. In Fig.~\ref{RLeo} we compare these observations to a fundamental-mode Mira model (D28847) at phase 0.2 (top) and to a first-overtone mode Mira model (E8560) at phase 0.21 (bottom). We have smoothed the visibility spectra here, using Eqn.~\ref{convoleqn}, assuming an intermediate FWHM of 45\,nm.  The upper and lower pairs of points in each panel are at spatial frequencies of $2$ and $3 \times 10^6$ rad$^{-1}$ respectively. Both model and observations show similar variations in visibility with wavelength at $2 \times 10^6$ rad$^{-1}$, although it is difficult to say more because of the large FWHM, the relative flatness of the visibility spectra here and the position of the 833\,nm filter being just outside the centre of the TiO band.

In summary, only one model shows a similar variation in visibility with wavelength to the observations. In most cases the models underestimate this variation and in several cases it is in the opposite sense to the observations, i.e. the TiO band (approx. 710\,nm) visibility is higher than the pseudo-continuum band (700\,nm) visibility in the models. However, this TiO band is subject to the extreme limb-darkening effect and the models may be unreliable here. This small sample leaves it unclear whether the models accurately predict the variation in visibility with wavelength for M giant stars. Our multiwavelength data for R~Dor and other stars will provide a better test of the accuracy of these models.

\section{A Model for Betelgeuse}
\begin{figure*}
\includegraphics[width=\the\hsize, totalheight=20cm]{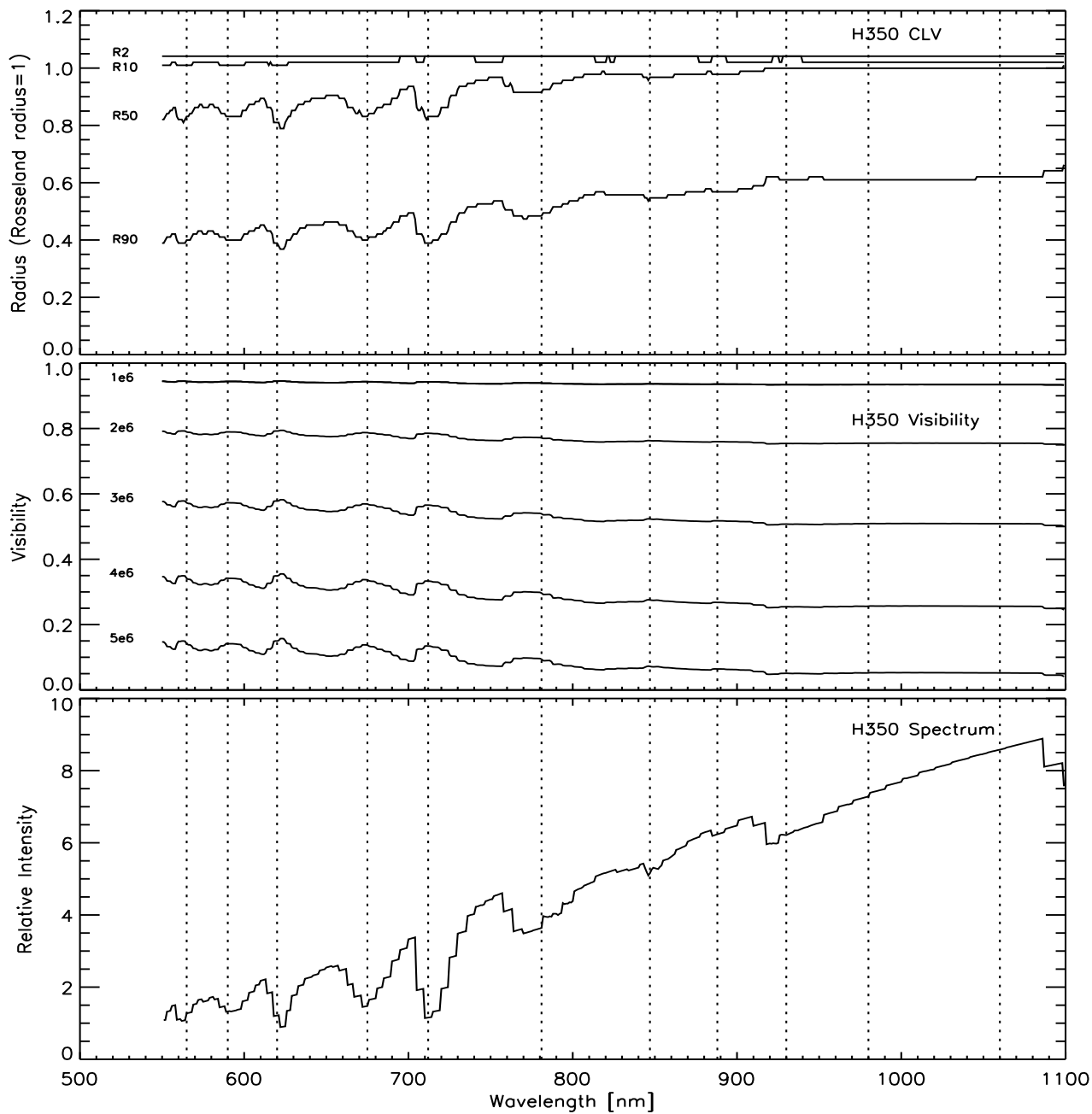}
\caption{As for Fig.~\ref{cvs} but for the 15M$_{\odot}$ static M giant model, H350. The flatness of the R2 and R10 CLV spectra is a consequence of the compactness of this model atmosphere. }
\label{cvsH350}
\end{figure*}

Fig.~\ref{cvsH350} shows the CLV, visibility and intensity spectra for the 15M$_{\odot}$ model (H350) which is used to represent the supergiant $\alpha$~Ori (Betelgeuse, M2 Iab). Unfortunately we have found no suitable observations (using the criterion in Section~\ref{opvs}) to compare with this model.

The model displays much smaller variations in both CLV and visibility with wavelength than later M giant models. In particular, the R2 radius is independent of wavelength, which reflects the fact that the H350 model atmosphere is nearly compact. Above about 900\,nm, its CLV and visibility spectra become constant. Below about 800\,nm, we once again see higher visibilities, within the first null, at TiO bands than in pseudo-continuum bands. A naive interpretation of this, without consideration of the detailed CLV shape, would be misleading. Therefore an observer of the H350 model star might conclude that the stellar radius is smaller at TiO than at pseudo-continuum wavelengths whereas, in reality, the atmosphere is compact and diameters are not wavelength-dependent. It is just that the limb-darkening effect is very pronounced in the TiO bands. In fact \citeone{CHH86} explained the scatter in previous wavelength-dependent angular diameter measurements of $\alpha$~Ori by adopting a varying limb-darkening behaviour. Although this effect appears similar to the extreme limb-darkening effect discussed in Section~\ref{anomolous}, we prefer to reserve that term for cool giants, in particular Miras, that have non-compact atmospheres. As discussed in that section, limb-darkening in the strongest TiO bands below 800\,nm for some of those stars may be so extreme that it increases the visibilities strongly within the first null and disguises the effects of geometric extension of the star's atmosphere. This peculiar situation is called the extreme limb-darkening effect in this paper.

\section{Conclusion}
Interferometric observations of M giants give us visibility profiles related to their centre-to-limb variations (CLVs). Observers never see the CLV directly and cannot measure an optical-depth radius (which is commonly used in theoretical work). They can only measure the visibility with a limited accuracy and to a limited spatial frequency which prevents reconstruction of the true CLV. Therefore observers convert standard limb-darkening profiles or CLVs from model photospheres of M giants to model visibility profiles and fit these to the observed visibility profiles. Only then can they define a radius, effective temperature and other parameters for the star. The MAPPIT interferometer is capable of recording multi-wavelength visibilities. Comparing these with the models should help improve the models and our understanding of M giants, particularly their temperature structure.

We have developed a way of displaying wavelength-dependent CLVs and visibilities. Our CLV and visibility spectra, as we have called them, show CLV versus wavelength for several radial positions and visibility versus wavelength for several spatial frequencies. The CLV spectra show both the overall shape and any low-level extension of the CLV at each wavelength while the visibility spectra allow direct comparisons with multi-wavelength observations. The CLV data are derived from the Mira and non-Mira M giant photospheric models of BBSW, BSW96 and HS98. The data are displayed at wavelengths from 550 to 1100\,nm but this could easily be extended to cover the range present in the models which is 320\,nm to 4100\,nm. We are also able to plot other radial positions or spatial frequencies and can easily and quickly simulate the effect of any real filter on the monochromatic CLV and visibility spectra. The visibility spectra provide a tool for comparing multi-wavelength interferometric observations of M giants with models of their centre-to-limb variation at a single epoch. We developed these plots to help interpret multi-wavelength interferometric observations that we have obtained with MAPPIT. The plots might also be useful when choosing filter properties (central wavelength, width and leakage) for narrow-band observations.

Care must be taken when making comparisons between interferometric observations and models. Several effects, including the model limitations and an extreme limb-darkening effect could lead to misinterpretations of the data. The extreme limb-darkening effect occurs in very strong TiO bands but only in some Mira models and some very cool non-Mira models. To an observer the model visibilities might suggest, at first glance, that the star appears smaller in the very strong absorption bands. Careful examination of high spatial frequency model-derived visibilities, or the model CLV, would show the star actually has a faint extended limb. The concept that an M giant will always appear larger, to an observer, in an absorption band may be too simple, at least when examining models. However, the few published observations that are relevant to this situation do not show this extreme effect, so further observations would be useful in establishing whether or not it is only a modelling artifact. Since a star's CLV depends on temperatures and temperature gradients, accurate interferometric observations of strong TiO bands would provide a sensitive method of probing the temperature structure of the outermost atmospheric layers and improving the models.

Our visibility spectra can now explain the similarity in visibilities of R~Dor observed by \citeone{BZvdL97} and \citeone{JBR97} despite the different filter properties used in those studies. We chose appropriate models from BBSW and BSW96 and found that the visibility spectrum is flat near 850\,nm and so a range of filter profiles will result in the same visibility profile.

At present there are only a limited number of observations that can be compared to models using our visibility spectra. In most cases the models show a much smaller variation in visibility with wavelength than the observations, and in some cases the models predict the reverse of the observed variation. Only one model, the fundamental-mode Mira (D28320), shows a similar variation in visibility with wavelength to the observations of R~Cas by \citeone{WBH96}. Our multi-wavelength observations of R~Dor, which are currently being processed, will provide a clearer test of how well the models predict the variation in visibility with wavelength for M giant stars.

\subsection*{Acknowledgements}
We are grateful to the Australian Research Council and The University of Sydney for financial support while carrying out this work.

\end{document}